%% file: revisedDOCFINAL2.tex
\documentclass[authoryear, review]{elsarticle} 

\usepackage{fullpage}

\usepackage{multirow}
\usepackage{comment}
\usepackage[latin1]{inputenc}
\usepackage{graphicx, pifont, geometry , amsmath, graphics,  amsfonts, amssymb, subfigure, amstext}
\usepackage{wrapfig, amsopn, latexsym, epsfig,multicol, verbatim}
\usepackage{amssymb,fancyhdr,  bm}

\usepackage[plainpages=false,pdfpagelabels,hypertexnames,linktocpage, colorlinks, citecolor=red,linkcolor=blue]{hyperref}

\begin{document}


\begin{frontmatter}


\title{A bootstrap test for equality of variances}
\author{Dexter O. Cahoy\corref{cor1}}
\ead{dcahoy@latech.edu}

\cortext[cor1]{Corresponding Author: Tel: +1 318 257 3529; Fax: +1 318 257 2182}
\address{Program of Mathematics and Statistics, College of Engineering and Science, Louisiana Tech University, Ruston, LA 71272}

\begin{abstract}
\indent \indent We introduce a bootstrap procedure to test the hypothesis $H_o$ that $K+1$ variances are homogeneous. The procedure uses a variance-based statistic, and is derived from a normal-theory test for equality of variances. The test equivalently expressed the hypothesis as $H_o: \bm{\eta}=( \eta_1,\ldots,\eta_{K+1})^T=\bm{0}$, where $\eta_i$'s are log contrasts of the population variances.  A box-type acceptance region is constructed to test the hypothesis $H_o$. Simulation results indicated that our method is generally superior to the Shoemaker and  Levene tests, and the bootstrapped version of Levene test in controlling the Type I and Type II errors.
\end{abstract}


\end{frontmatter}

\input{equa1b.tex}

\input{equa2b.tex}

\input{equa3b.tex}

\section*{Acknowledgment}
\emph{The author is  grateful to the  editors especially the two reviewers for their insightful comments and suggestions that significantly improved the article}. 

\section*{References}







\end{document}

%% file: equa1b.tex
\section{Introduction} \label{a}
Testing the homogeneity of variances arises in many scientific applications. It is increasingly used now  to determine uniformity in quality control, in biology, in agricultural production systems, and even in the development of educational methods \citep[see][]{bbrow04}. It is also a prelude to testing the equality of population means such as the analysis of variance (ANOVA)\citep[see][]{sche59}, dose-
response modeling or discriminant analysis. The literature for testing equality of variances is huge and we refer the readers to  the comprehensive review of \cite{cojj81}.

More recently, procedures for testing equality of variances that are robust to non-normality have been categorized into three major approaches. These strategies are based on the following: (1) Kurtosis adjustment of normal-theory tests \citep{boxa55,shoe03},  (2) Analysis of variance (ANOVA) on scale variables such as the absolute deviations from the median or mean \citep{levene60, baf74}, and  (3) Resampling methods  to obtain p-values for a given
test statistic \citep{boxa55,bbrow89}.  Descriptions of these methods are summarized  in \cite{bbrow04}.

The main focus of our research is on resampling methods as  they have been  shown to improve the Type I and possibly the Type II error rates  \citep[see][]{bbrow89,limlow96}. More specifically, our goal is to propose a variance-based procedure to test the homoscedasticity of variances for a wide variety of distributions.  It is also our objective to validate whether resampling methods improve Type I and Type II error rates.  An important attribute of our proposed method is its ability to control better the Type I and Type II error rates for small sample (both equal and unequal) sizes. Our test uses a box-type acceptance region rather than a p-value which distinguishes it from other resampling methods.  It is solely based on a variance-based statistic without applying any transformation to the observed data like smoothing, fractional trimming, or replacing original observations by the scale or residual estimates.  The variance-based procedure is also shown to be more sensitive to  deviations from the null conditions.

Just like \cite{bbrow89}, we prefer variance-based procedures as they are more appealing to practitioners, easier to interpret, and variances are of interest in many areas. We also hope that with constantly improving state-of-the-art computing machinery,  this research will encourage the use of resampling-based tests for equality of variances by practitioners, and  the  integration of these procedures into major statistical software packages.  The descriptions of the  bootstrap and non-bootstrap tests for equality of variances are given in Section 2. Section 3 shows the small-to-moderate sample size performance of the tests. We close the article with a  summary and an outline of possible future extensions.

%% file: equa2b.tex
\section{Description Of Tests To Be Compared}

Given $K+1$ samples  from the populations $F\lbrace (d_i-\mu_i)/\sigma_i \rbrace, i=1,\ldots,K+1$ with equal kurtosis, the $i$th sample $d_{i1}, d_{i2}, \ldots, d_{in_i}$ having size $n_i$, and $n=n_1+\ldots+n_{K+1}$, consider a test of the hypothesis
\[
H_o: \sigma_1^2=\sigma_2^2=\cdots=\sigma_{K+1}^2
\]
against the alternative hypothesis $H_a$ that at least two of the $K+1$ variances are unequal. Let $s_i^2$ denote the sample variance based on $n_i$ observations from the $i$th sample. We now describe the tests  that will be  compared.

\subsection{Levene's Test $L$}

\cite{levene60} first proposed ANOVA on the scale variables  $e_{ij}=|d_{ij}-\overline{d}_i|,$ where $\overline{d}_i$ is the mean of the $i$th sample but \cite{mil68} showed that $e_{ij}$ is asymptotically correct for asymmetric populations if the median is used instead of the mean. \cite{baf74}  formally
studied Levene's method where the median was used instead of the mean to center the variables.  \cite{bbrow89} and \cite{limlow96} provide more details on the features of Levene's test.

We consider Levene's test as it is widely used in practice even if it is not a variance- and resampling-based procedure. It is also recommended by \cite{cojj81}. Levene's procedure is a test for equality of means applied to the scale quantities  $e_{ij}=|d_{ij}-\tilde{\mu}_i|,$ where $\tilde{\mu}_i$ is the median of the $i$th sample $\lbrace d_{ij}, j=1,\ldots, n_i \rbrace$. The test statistic is
\[
L_{ts}=\frac{\sum\limits_{i=1}^{K+1}n_i(e_{i\cdot}-\overline{e}_{\cdot \cdot})^2/K}{\sum\limits_{i=1}^{K+1}\sum\limits_{j=1}^{n_i}(e_{ij}-\overline{e}_{i \cdot})^2/(n-K+1)},
\]
where $e_{i\cdot}=\sum\limits_{j=1}^{n_i} e_{ij}/n_i $ and $\overline{e}_{\cdot \cdot}=\sum\limits_{i=1}^{K+1}\sum\limits_{j=1}^{n_i}e_{ij}/n$. We reject the null hypothesis $H_o$ if $L_{ts}$ exceeds the $100(1-\alpha)$th quantile $F_{K,n-K+1} (\alpha)$ of the $F$-distribution with $K$ and $n-K+1$ degrees of freedom.

Some variants of the Levene's test are proposed by \cite{obri78}, \cite{hah00}, and \cite{goo00}.  But these modified versions of  Levene's test are still inferior in terms of level and power to its bootstrap version \citep[see][]{liu06},  which is discussed in Section 1.

\subsection{Shoemaker's Test $S$}
We also consider Shoemaker's test $S$, which not only provides good insights to our procedure but was the test recommended by \cite{shoe03}   after comparing its performance with some kurtosis-adjusted normal-theory tests.  The test statistic is
\[
S_{ts}=\sum\limits_{i=1}^{K+1} \Big(\ln s_i^2 - \overline{\ln s^2} \Big)^2/ \widehat{\text{var}} (\ln s_i^2),
\]
where $\overline{\ln s^2}=\sum\limits_{i=1}^{K+1} \ln s_i^2/(K+1)$, $\widehat{\text{var}} (\ln s_i^2)= \big[ \widehat{\mu_4^{'}}/\widehat{\sigma}^4 - (h_i-3)/(h_i-1) \big]\big/h_i$, $h_i$ is the harmonic mean $(K+1)/\sum\limits_{i=1}^{K+1} 1/n_i$,  $\widehat{\mu_4^{'}}=\sum\limits_{i=1}^{K+1}\sum\limits_{j=1}^{n_i} (d_{ij}-\overline{d}_{i\cdot})^4/n$ is the estimator of the fourth moment about the population mean, and $\widehat{\sigma}^2= \sum\limits_{i=1}^{K+1} (n_i-1)s_i^2/n$. He recommended the estimator of an asymptotically equivalent formula which is $\widehat{\text{var}} (\ln s_i^2)=\big[ \widehat{\mu_4^{'}}/\widehat{\sigma}^4 - (h_i-3)/h_i \big]/(h_i-1)$ to improve simulation accuracy. The null hypothesis is rejected when $S_{ts}$ exceeds the $100(1-\alpha)$th percentile of the chi-square distribution with $K$ degrees of freedom.

\subsection{Lim and Loh's Test $BL$}

\cite{limlow96} compared several bootstrap and non-bootstrap tests for heterogeneity of variances. A bootstrap version of the Levene's test was recommended because of its superiority in terms of  power and Type I error robustness. The procedure used the technique of \cite{bbrow89}, and is given below.

1.) Compute the test statistic  from the given data $d_{ij}, i=1,\ldots,K+1, j=1,\ldots, n_i$.

2.) Initialize $l=0$.

3.) Compute the residuals $e_{ij}=d_{ij}-\widehat{\mu}_i, i=1,\ldots,K+1, j=1,\ldots,n_i$ where $\widehat{\mu}_i$ is the median of  group $i$.

4.) Draw $n$ data points $e_{ij}^*$'s from the pooled residuals $\overline{R}=\lbrace d_{ij}-\widehat{\mu}_i, i=1,\ldots,K+1, j=1,\ldots,n_i\rbrace$.

5.) If the sample size $n_i$  of the $i$th group is  less than 10 then smooth the bootstrap observations by setting $d_{ij}^*=(12/13)^{0.5} ( e_{ij}^*+ qU)$, where $q^2=\sum\sum (d_{ij}-\overline{d}_i)^2/n$ and $U$ is an independent and uniformly distributed random variable on $(-1/2, 1/2)$. Otherwise, set $d_{ij}^*=e_{ij}^*$.

6.) Compute the test statistic value $L_{ts}^*$ based on the bootstrapped samples $d_{ij}^*$. If $L_{ts}^*>L_{ts}$ then $l=l+1$.

7.) Repeat steps 4, 5, 6, $B$ times.

8.) The bootstrap p-value is $l/B$.

\cite{limlow96}'s test  $BL$ rejects the hypothesis  $H_o$ if the p-value $l/B < \alpha$. This procedure is also recommended by \cite{bbrow04} as it provides more power and improves the Type I error robustness.

Note that the bootstrap version of the Bartlett's test is another alternative especially when the populations do not have large kurtosis. However, we exclude it in the comparison as it is not the recommended procedure by the previous studies of \cite{limlow96} and  \cite{bbrow04}.  More importantly, it is not ``robust" (unlike the proposed test)  for  highly leptokurtic distributions, which can be tricky in practice for small and/or unequal sample sizes.  However, \cite{bbrow89, bbrow04}  recommended the bootstrap version Bartlett's test for comparing larger number of populations.  A comparison of our proposed method with the bootstrap version of the Bartlett's test especially involving large number of groups would be an  interesting extension of our study as well.

\subsection{Alam and Cahoy's Test $T$}

We now give a brief background on how we derive our variance-based test statistic. \cite{alca99} proposed the following test for equality of variances for normal populations:  Let $\pi_1,\ldots,\pi_{K+1}$ be $K+1$ normal populations, and let $\mu_i, \sigma_i^2$ be the mean and variance of $\pi_i, i=1,\ldots,K+1$. Let
$z_i=(n_i-1)s_i^2/s^2, i=1,\ldots,K+1,$ where $s^2=\sum\limits_{j=1}^{K+1} (n_j-1) s_j^2.$ Under $H_o$, it follows that $z = (z_1,\ldots,z_{K+1})^T$ is jointly distributed according to the Dirichlet distribution $D_K (z;\nu_1,\ldots,\nu_{K+1})$ (see \cite{balnev03} ),  and is given by
\[
D_K(\bm{x}; \nu_1,\ldots,\nu_{K+1})=\frac{\Gamma(\nu)}{\prod\limits_{j=1}^{K+1} \Gamma(\nu_j)}x_1^{\nu_1-1}\cdots x_K^{\nu_K-1}\Bigg(1-\sum\limits_{j=1}^K x_j \Bigg)^{\nu_{K+1}-1}, \label{1e1}
\]
where $\bm{x}=(x_1,\ldots,x_{K+1})^T, x_i \geq 0, \nu_i=(n_i-1)/2 >0, i=1,\ldots, K+1, x_{K+1}=1-\sum\limits_{j=1}^K x_j,$ and  $\nu=\nu_1+\cdots + \nu_{K+1}$.
Let
\[
\eta_i=\ln \Bigg( \frac{\sigma_i^2}{\big(\prod\limits_{j=1}^{K+1}\sigma_j^2\big)^\frac{1}{K+1}}\Bigg),\qquad i=1,\ldots,K+1.
\]
The null hypothesis $H_o$ can then be expressed as   $\bm{\eta}=( \eta_1,\ldots,\eta_{K+1})^T=\bm{0}$, versus $\eta_i \neq 0$ for at least one value of $i$, under the alternative hypothesis $H_a$. We construct a box-type confidence region for $\bm{\eta}$ as follows: Let $\bm{y}=(y_1,\ldots, y_{K+1})^T$ and $y_i$ be given by
\begin{align}
y_i & = \ln z_i -\frac{1}{K+1} \sum\limits_{j=1}^{K+1} \ln z_i \notag \\
&=\ln \Big(s_i^2 \; / \;\big(\prod\limits_{j=1}^{K+1}s_j^2\big)^\frac{1}{K+1} \Big) + \nu_i^{'}, \notag
\end{align}
where $\nu_i^{'}=\ln \Big(\nu_i \; / \;\big(\prod\limits_{j=1}^{K+1}\nu_j\big)^\frac{1}{K+1} \Big), i=1, \ldots,K+1.$  Let $\bm{w}=(w_1,\ldots, w_{K+1})^T$ and $ w_i  = \ln x_i - \sum\limits_{j=1}^{K+1} \ln x_j/(K+1), \linebreak  i=1, \ldots,K+1.$ It then follows that $ y_i \stackrel{d}{\sim}w_i+\eta_i, i=1,\ldots, K+1$. Moreover, let $\theta_i$ and $\lambda_i^2$ be the mean and variance of $w_i$. A $(1-\alpha)$-level confidence region for $\bm{\eta}$ is given by
\begin{equation}
C=\lbrace \bm{\eta}: y_i-\theta_i-c \lambda_i \leq \eta_i \leq y_i- \theta_i+c \lambda_i, \; i=1,\ldots, K+1\rbrace, \label{1e8}
\end{equation}
where $c>0$ is chosen such that
\begin{equation}
\mathbf{P}\lbrace\bm{w}: \theta_i-c \lambda_i \leq w_i \leq \theta_i+c \lambda_i, \; i=1,\ldots, K+1\rbrace=1-\alpha. \label{1e9}
\end{equation}
The test $T$ of the hypothesis $H_o$ (of level $\alpha$) is derived from (\ref{1e8}) with the acceptance region given by
\[
A=\lbrace\bm{y}: \theta_i-c \lambda_i \leq y_i \leq \theta_i+c \lambda_i, \; i=1,\ldots, K+1\rbrace. \label{1e10}
\]
The value of $c$ is calculated numerically from (\ref{1e9}) using the distribution of $\bm{w}$ which is
\[
g(\bm{w})=\frac{(K+1)\Gamma(\nu)}{\prod\limits_{j=1}^{K+1}\Gamma(\nu_j)}\bigg( \sum\limits_{j=1}^{K+1}\exp (w_j) \bigg)^{-\nu} \exp\bigg(\sum\limits_{j=1}^{K+1}\nu_j w_j \bigg), \label{2.1}
\]
where $ -\infty < w_i < \infty,$ and $\sum\limits_{j=1}^{K+1} w_j=0$. This test statistic is powerfully sensitive to individual deviations in the values of the $\eta_i$'s from the origin, under the alternative hypothesis $H_a$. \cite{alca99} give the moments of $\bm{w}$, its asymptotic properties, and the critical value $c$ for normal populations.

We now construct the bootstrap version of the above test. For any distribution and with a slight  modification, a generalized box-type confidence region is now given by
\[
C^{'}=\lbrace\bm{\eta}:  \widehat{\eta}_i-c^{'} \lambda_i^{'}  \leq \eta_i \leq \widehat{\eta}_i + c^{'} \lambda_i^{'}, \; i=1,\ldots, K+1\rbrace,  
\]
where $\mathbf{E}(\widehat{\eta}_i)=\eta_i$, and  $\lambda_i^{'}= \text{var} (\widehat{\eta}_i)^{1/2}$.  Using the variance stabilizing transformation of the sample variance, the  mean  and variance of $\ln s_i^2$ can be approximated by $\mathbf{E} (\ln s_i^2)=\ln \sigma_i^2$, and  $\text{var} (\ln s_i^2)=\big[ \mu_4^{'}/\sigma^4 - (n_i-3)/(n_i-1) \big]\big/n_i$ for any distribution. But just like \cite{shoe03}, we use the asymptotically equivalent formula $\text{var} (\ln s_i^2)=\big[ \mu_4^{'}/\sigma^4 - (n_i-3)/n_i \big]\big/(n_i-1)$ except that we don't use the harmonic mean for the sample size $n_i$. When the null hypothesis $H_o$ is true, the box-type acceptance region of our test for any distribution can be approximated by
\[
\widehat{A}=\lbrace\bm{t}: -c^{'}  \leq t_i \leq c^{'}, \; i=1,\ldots, K+1\rbrace,  \label{1e12}
\]
where $\bm{t}= (t_1, \ldots, t_{K+1})^T, \eta_i=0, t_i =\widehat{\eta}_i/\widehat{\lambda}_i$ is the test statistic,  $c^{'}$ is the critical value that needs to be found such that $\widehat{A}$ has  the coverage  $\mathbf{P}(\widehat{A})=1-\alpha,  \widehat{\eta}_i= \ln \big(s_i^2 \; / \;\big(\prod\limits_{j=1}^{K+1}s_j^2\big)^{1/(K+1)} \big),$  and $\widehat{\lambda}_i= \Big\lbrace\big[1-2/(K+1)\big]\widehat{\text{var}} (\ln s_i^2) + (1/(K+1)^2)\sum\limits_{j=1}^{K+1} \widehat{\text{var}} (\ln s_j^2 )\Big\rbrace^{1/2}.$ Consequently, the bootstrap version of the box-type acceptance region is then given by
\[
\widehat{A^{*}}=\lbrace\bm{t^{*}}: -c^{*} \leq t_i^{*} - \widehat{t}_i \leq c^{*}, \; i=1,\ldots, K+1\rbrace,  \label{1e12}
\]
where $\bm{t}^*= (t_1^*, \ldots, t_{K+1}^*)^T,  t_i^{*}=\widehat{\eta_i^*}/ \widehat{\lambda_i^*}, \widehat{t}_i = \overline{t_i^{*}}=\sum\limits_{i=1}^B t_i^{*} / B$,  $\widehat{\lambda_i^*}= \Big\lbrace\big[1-2/(K+1)\big] \widehat{\text{var}} (\ln s_i^{2^*}) + (1/(K+1)^2)\sum\limits_{j=1}^{K+1} \widehat{\text{var}} (\ln s_j^{2^*} )\Big\rbrace^{1/2},$ and $\widehat{\text{var}} (\ln s_i^{2^*})=\big[ \widehat{\mu}_4^*/\widehat{\sigma}^{4^*} - (n_i-3)/n_i \big]\big/(n_i-1)$. The test now is being reduced to finding the critical value $c^{*}$. Note that the availability of the standard error estimate without necessarily performing a second layer bootstrap makes the calculations faster. We emphasize that a viable alternative is to use the pivotal quantity $(\widehat{\eta_i^*}-\widehat{\eta_i})/ \widehat{\lambda_i^*}$. But it often gives more conservative estimated sizes  and  smaller power (than our procedure but is still better than the other tests in controlling both the Type I and Type II errors). 
Below is the algorithm for a given $\alpha$: 

1.) Calculate the test statistic value  $t_i= \widehat{\eta}_i/\widehat{\lambda}_i$ from the observed data $\lbrace d_{ij}, j=1,\ldots, n_i\rbrace, i=1,\ldots,K+1$.

2.) Draw $n_i$ data points $d_{ij}^*$'s with replacement from each  sample  $\lbrace d_{ij}, j=1,\ldots,n_i\rbrace, i=1,\ldots,K+1$.

3.) Compute the bootstrap test  statistic $t_i^{*}=\widehat{\eta}_i^{*}/ \widehat{\lambda}_i^{*}, i=1, \ldots, K+1$.

4.) Repeat steps 2 and 3, $B$ times.

5.) Center  $t_i^{*}$'s by subtracting the $i$th bootstrap mean, i.e., let $t_i^{**}  = t_i^* - \overline{t_i^*},  i=1,\ldots, K+1$.

6.) Sort all the centered $t_i^*$'s ($B(K+1)$ of them) in descending order as $t^{**(1)} \geq t^{**(2)}\geq t^{**(3)}\geq \ldots \geq t^{**(B(K+1))}$.

7.) For $l=1,\ldots, B(K+1)$, if $\frac{\# \big\lbrace    -t^{**(l)} \leq t_1^{**} \leq  t^{**(l)},   -t^{**(l)} \leq t_2^{**} \leq t^{**(l)}, \ldots, -t^{**(l)} \leq t_{K+1}^{**} \leq t^{**(l)} \big\rbrace }{B} = 1-\alpha$ then stop, and the critical value is given by $c^*=t^{**(l)}$.

The test rejects $H_o$ if $|t_i|=\big|\widehat{\eta}_i / \widehat{\lambda}_i \big|>c^*$ for at least one $i, i=1, \ldots, K+1$.  Notice also that we choose the box-type confidence or acceptance region centered at the origin, where the boundaries are parallel to the axes and have equal lengths.   We are still studying how to efficiently calculate these critical values for rectangular prisms having unequal lengths or for likelihood-based regions as in \cite{hal87}.

%% file: equa3b.tex
\section{Empirical Results} \label{g}

In our simulation study, we compared the Type I error robustness  of the tests using 36 sample size-distribution combinations.   The power was examined using 5 and 6 variance configurations for equal- and unequal-sample cases, respectively. In addition, we considered 6 small-to-moderate sample size configurations.  Six distributions with kurtosis $\kappa$ ranging from 1.8 to 9 were selected.  These distributions are as follows: (\emph{i}) uniform ($\kappa=1.8$), (\emph{ii}) Gaussian ($\kappa=3$), (\emph{iii}) extreme value ($\kappa=5.4$), (\emph{iv}) Laplace ($\kappa=6$), (\emph{v}) Student's $t_5$ with 5 degrees of freedom ($\kappa=9$), and (\emph{vi}) exponential ($\kappa=9$). This array of distributions was considered by \cite{bbrow89} and \cite{limlow96} to be representative of the data encountered in practice.  The extreme value has the probability density function $f(x) =\exp (-x) \exp(-\exp(-x))$ \citep[see][]{sc01}. All variances under the null were chosen to be one.  The estimated power and significance levels of the tests were compared for two-sample,  three-sample, and four-sample cases. The simulations used the random number generator ``Mersenne-Twister" which is a twisted generalized feedback shift register (GFSR) with period $2^{19937} - 1$ and is equidistributed in 623 consecutive dimensions over the whole period \citep[see][]{matni98}.

Following \cite{bbrow89}, we performed 1000 Monte Carlo simulations using $B=500$ bootstrap samples for each run.  We adopted \cite{cojj81}'s criterion to assess Type I error robustness. It said that a test is ``robust"  if the maximum estimated significance level over all the sample size-distribution null combinations (equal variances) is less than twice the nominal level. We used the nominal level $\alpha=0.05$. We highlighted estimated levels that  exceeded 0.10 using an asterisk.

\subsection{Two-Sample Case  $(K=1)$}

In this case, the null conditions included the 6 sample size combinations $n_1=n_2=5,10,15,$ $(n_1, n_2)=(7, 10)$, $(7, 15)$, and $(10, 15)$. We also considered the variance ratios $(\sigma_1^2, \sigma_2^2) = (1,10)$, $(1,16)$, and $(16,1)$.    Table 1 shows the estimated sizes of the tests. It clearly indicates that  all the tests except \cite{shoe03}'s test $S$ are robust according to \cite{cojj81}'s criterion. \cite{shoe03}'s test has a large maximum estimated size of 0.13 which corresponds to the sample size combination $(n_1,n_2)=(5, 5)$ under the exponential distribution. In addition, the test $S$ seemed to be sensitive to the sample size configurations as shown by the inflated Type I error rates for unequal sample sizes. Meanwhile, our test $T$ has a maximum test size estimate of 0.08, while \cite{levene60}'s $L$ and \cite{limlow96}'s $BL$,  have 0.04 and  0.06,  respectively. This confirmed the previous observations of \cite{cojj81}, \cite{bbrow89}, and \cite{limlow96} about the extreme conservativeness of the Levene's test $L$.  These results also imply that our test $T$ controls the Type 1 error better than Levene's test $L$ and is less conservative than the bootstrap Levene's test $BL$. This observation is even more noticeable in the case of having unequal sample sizes. 

Tables 2 and 3 show the simulated power of the tests. From here on, we excluded \cite{shoe03}'s test $S$ as it was not ``robust" under \cite{cojj81}'s criterion over the 36 prescribed  null settings. The variance ratios under the alternative hypothesis are chosen to minimize unity in power across all the distributions for moderate sample sizes. For equal sample sizes, the alternative hypothesis has the variance configuration $(\sigma_1^2, \sigma_2^2) = (1,10)$.  It is apparent that our test $T$ has the highest power averaged over all the distributions.  With an average power of $26\%$  for sample size configuration $n_1=n_2=5$, it is more than thrice the power of \cite{levene60}'s test which is $8\%$, but is just slightly greater than that of \cite{limlow96} $BL$'s $25\%$.  Furthermore, the superiority of our test becomes more noticeable  when the sample size reaches $n_1=n_2=7$ and 10. However, the power difference becomes  negligible when the sample size exceeds $n_1=n_2=15$. Moreover, both Levene's $L$ and its bootstrap version $BL$ tend to approach unity faster as the sample size increases under the exponential distribution.

As noticed by \cite{low87} and \cite{limlow96}, the power of \cite{limlow96}'s  test $BL$  and \cite{levene60}'s $L$ is low when the large $n_i$ are associated with large $\sigma_i^2$, and is high if large $n_i$ is associated with small $\sigma_i^2$. This led us to average the power of  the tests corresponding to variance configurations $(\sigma_1^2, \sigma_2^2) = (1,16)$ and $(16,1)$  for unequal sample sizes, and is shown in Table 3.  From the table, it is clear that our test $T$ still dominated the other procedures across all the distributions.  More specifically,  the averaged power of our test $T$ could possibly be at least $10\%$ higher than the Levene's test $L$ but is just slightly more powerful than the bootstrap Levene's test $BL$.

Overall, our procedure stood out to be the most powerful and is the least conservative test among all other ``robust"  procedures for the two-sample case. Our results also confirmed that bootstrapping Levene's test  $L$ corrected the conservativeness of its Type I error rate and improved its power.

\renewcommand{\arraystretch}{1.0}
\begin{table}[h!t!b!p!]
\caption{\emph{Estimated sizes for testing  $H_o:\sigma_1^2=\sigma_2^2$ at level 0.05,  for  different sample size combinations. The Monte Carlo estimates are based on 1000 replications, and the standard error of the entries is bounded by 0.016.}} \centerline {
\begin{tabular*}{5.4in}{@{\extracolsep{\fill}}c||c@{\hspace{0.01in}}c@{\hspace{0.01in}}c@{\hspace{0.01in}}c@{\hspace{0.01in}}c@{\hspace{0.01in}}c@{\hspace{0.01in}}}
\hline
Test& Uniform & Normal & Extreme &   Laplace	& Student's $t_5$	& Exponential\\
\hline \hline
& \multicolumn{6}{c}{$n_1=n_2=5$} \\
$T$ &	0.02 &	0.04	&0.04	&0.05	&0.05	&0.04 \\
$L$ &   0.00 &	0.01	&0.01	&0.01	&0.01	&0.01\\
$BL$&   0.03 &	0.04	&0.04	&0.04	&0.05	&0.06\\
$S$	&   0.04 & 0.05	&0.07	&0.09	&0.08	&$0.13^\textbf{*}$\\
& \multicolumn{6}{c}{$n_1=n_2=10$} \\
$T$ & 0.05 &	0.06 	&0.07	&0.07	&0.07	&0.07 \\
$L$	& 0.04	 &   0.03	&0.04	&0.04	&0.04	&0.04\\
$BL$& 0.05	&0.04	&0.05	&0.05	&0.05	&0.05 \\
$S$	&0.03	&0.06	&0.07	&0.08	&0.06	&$0.12^\textbf{*}$ \\
& \multicolumn{6}{c}{$n_1=n_2=15$} \\
$T$	&0.05	&0.05	&0.06	&0.06	&0.07	&0.07\\
$L$	&0.02	&0.03	&0.03	&0.04	&0.02	&0.04\\
$BL$&0.05	&0.04	&0.06	&0.05	&0.04	&0.05\\
$S$	&0.03	&0.06	&0.08	&0.07	&0.07	&0.10\\
\hline
& \multicolumn{6}{c}{$n_1=5, n_2=10$} \\
$T$	&0.03	&0.05	&0.04	&0.08	&0.06	&0.06\\
$L$	&0.03	&0.03	&0.02	&0.02	&0.03	&0.04\\
$BL$	&0.04	&0.05	&0.04	&0.05	&0.06	&0.06\\
$S$	& 0.06	&0.06 	&0.08	&$0.11^\textbf{*}$ & 0.08	&$0.12^\textbf{*}$ \\
& \multicolumn{6}{c}{$n_1=7, n_2=15$} \\	
$T$	&0.05	&0.07	&0.07	&0.07	&0.07	&0.07\\
$L$	&0.02	&0.04	&0.02	&0.03	&0.03	&0.04\\
$BL$	&0.05	&0.06	&0.04	&0.05	&0.05	&0.06\\
$S$	& 0.05	&0.04	&0.07	&0.09	&0.06	&$0.11^\textbf{*}$\\
& \multicolumn{6}{c}{$n_1=10, n_2=15$} \\
$T$	&0.05	&0.06	&0.07	&0.07	&0.08	&0.07\\
$L$	&0.04	&0.03	&0.04	&0.03	&0.03	&0.04\\
$BL$&0.06	&0.04	&0.06	&0.04	&0.05	&0.05\\
$S$	& 0.04	&0.06	&0.08	&0.07	&0.06	&$0.11^\textbf{*}$\\
\hline
\end{tabular*}
}
\footnotemark{Note: $^\textbf{*}$ indicates significantly higher than twice the nominal level $\alpha=0.05$}
\end{table}

\begin{table}[h!t!b!p!]
\caption{\emph{Estimated power of the tests at level 0.05 for equal sample sizes. The Monte Carlo estimates are based on 1000 replications, and the standard error of the entries is bounded by 0.016. }} \centerline {
\begin{tabular*}{5.4in}{@{\extracolsep{\fill}}c||c@{\hspace{0.01in}}c@{\hspace{0.01in}}c@{\hspace{0.01in}}c@{\hspace{0.01in}}c@{\hspace{0.01in}}c@{\hspace{0.01in}}|c@{\hspace{0.01in}}}
\hline
Test& Uniform & Normal & Extreme &   Laplace	& Student's $t_5$	& Exponential & Average\\
\hline \hline
& \multicolumn{6}{c|}{$(\sigma_1^2, \sigma_2^2)=(1,10)$} \\
\hline 	
& \multicolumn{6}{c|}{$n_1=n_2=5$} \\
$T$ &0.28	&0.29	&0.27	&0.27	&0.25	&0.22	&0.26\\
$L$ &0.10	&0.08	&0.08	&0.08	&0.07	&0.09	&0.08\\
$BL$ &0.30	&0.29	&0.25	&0.23	&0.23	&0.22	&0.25\\
& \multicolumn{6}{c|}{$n_1=n_2=7$} \\
$T$ & 0.70	&0.58	&0.48	&0.41	&0.48	&0.33	&0.50\\
$L$ &0.39	&0.32	&0.25	&0.21	&0.27	&0.21	&0.27\\
$BL$ &0.55	&0.48	&0.40	&0.36	&0.42	&0.32	&0.42\\
& \multicolumn{6}{c|}{$n_1=n_2=10$} \\
$T$ &0.94	&0.82	&0.66	&0.58	&0.69	&0.42	&0.69\\
$L$	&0.78	&0.67	&0.56	&0.47	&0.58	&0.38	&0.58\\
$BL$&0.81	&0.72	&0.61	&0.53	&0.63	&0.42	&0.62\\
& \multicolumn{6}{c|}{$n_1=n_2=15$} \\
$T$	&1.00	&0.96	&0.87	&0.78	&0.85	&0.61	&0.84\\
$L$	&0.97	&0.91	&0.83	&0.76	&0.81	&0.63	&0.82\\
$BL$&0.98	&0.93	&0.86	&0.79	&0.85	&0.67	&0.85\\
\hline 	
\end{tabular*}
}
\end{table}

\begin{table}[h!t!b!p!]
\caption{\emph{Estimated power at level 0.05 averaged over variance ratios $(\sigma_1^2, \sigma_2^2)= (1, 16)$ and $(16, 1)$  for unequal sample sizes. The Monte Carlo estimates are based on 1000 replications, and the standard error of the entries is bounded by 0.016.}} \centerline {
\begin{tabular*}{5.4in}{@{\extracolsep{\fill}}c||c@{\hspace{0.01in}}c@{\hspace{0.01in}}c@{\hspace{0.01in}}c@{\hspace{0.01in}}c@{\hspace{0.01in}}c@{\hspace{0.01in}}|c@{\hspace{0.01in}}}
\hline
Test& Uniform & Normal & Extreme &   Laplace	& Student's $t_5$	& Exponential & Average\\
\hline \hline
& \multicolumn{6}{c|}{$n_1=5,  n_2=10$} \\
$T$ &0.69	 &0.61	 &0.52	 &0.47	 &0.54	 &0.41	 &0.54\\
$L$ &0.55	 &0.45	 &0.39	 &0.33	 &0.38	 &0.31	 &0.40\\
$BL$ &0.68	 &0.58	 &0.51	 &0.44	 &0.52	 &0.41	 &0.52\\
& \multicolumn{6}{c|}{$n_1=7, n_2=15$} \\
$T$ &0.94	 &0.70	 &0.77	 &0.69	 &0.77	 &0.55	 &0.74\\
$L$	&0.88	 &0.58	 &0.66	 &0.56	 &0.66	 &0.48	 &0.64\\
$BL$	&0.93	 &0.65	 &0.74	 &0.65	 &0.74	 &0.55	 &0.71\\
& \multicolumn{6}{c|}{$n_1=10, n_2=15$} \\
$T$ &0.99	 &0.97	 &0.88  &0.79	 &0.88	 &0.65	 &0.89\\
$L$	&0.97	 &0.92	 &0.82	 &0.73	 &0.84	 &0.63	 &0.82\\
$BL$&0.98	 &0.94	 &0.87	 &0.78	 &0.88	 &0.68	 &0.85\\
\hline
\end{tabular*}
}
\end{table}

\subsection{Three-Sample  Case $(K=2)$ }

Table 4 gives the estimated levels of the three tests for the 6 sample size combinations $n_1=n_2=n_3=7,10,15,$  $(n_1,n_2,n_3)=(7, 10, 15)$, $(7, 10, 20)$, and $(10, 15, 15)$. The table suggests that  all the three tests  are robust according to \cite{cojj81}'s criterion. The test  $T$ has a maximum size of 0.075 while \cite{levene60}'s $L$ and \cite{limlow96}'s $BL$ have maxima 0.05 and 0.06, correspondingly. The table also indicates that our procedure seems to be more conservative than the bootstrap Levene's test $BL$ for distributions with smaller kurtosis (e.g., uniform distribution) and with relatively small sample sizes (e.g., $n_1=n_2=n_3=7$). With unequal sample sizes, our procedure is the least conservative procedure except for the sample size combination $(n_1, n_2, n_3)=(7, 10, 15)$ under the uniform distribution.

\begin{table}[h!t!b!p!]
\caption{\emph{Estimated sizes for testing  $H_o:\sigma_1^2=\sigma_2^2=\sigma_3^2$ at level 0.05, for  different sample size combinations. The Monte Carlo estimates are based on 1000 replications, and the standard error of the entries is bounded by 0.016.}} \centerline {
\begin{tabular*}{5.4in}{@{\extracolsep{\fill}}c||c@{\hspace{0.01in}}c@{\hspace{0.01in}}c@{\hspace{0.01in}}c@{\hspace{0.01in}}c@{\hspace{0.01in}}c@{\hspace{0.01in}}}
\hline
Test& Uniform & Normal & Extreme &   Laplace	& Student's $t_5$	& Exponential\\
\hline \hline	
& \multicolumn{6}{c}{$n_1=n_2=n_3=7$} \\ 	
$T$ &0.03	&0.03	&0.03	&0.06	&0.06	&0.05\\
$L$	&0.01	&0.01	&0.02	&0.02	&0.02	&0.03\\
$BL$&0.03	&0.04	&0.05	&0.05	&0.06	&0.06\\
& \multicolumn{6}{c}{$n_1=n_2=n_3=10$} \\ 	
$T$ &0.03	&0.06	&0.06	&0.07	&0.07	&0.08\\
$L$	&0.03	&0.03	&0.04	&0.02	&0.04	&0.05\\
$BL$&0.04	&0.05	&0.04	&0.05	&0.06	&0.05\\
& \multicolumn{6}{c}{$n_1=n_2=n_3=15$} \\ 	
$T$	&0.04	&0.07	&0.07	&0.07	&0.06	&0.07\\
$L$	&0.02	&0.03	&0.04	&0.04	&0.03	&0.05\\
$BL$&0.04	&0.05	&0.06	&0.05	&0.05	&0.06\\
\hline
& \multicolumn{6}{c}{$n_1=7, n_2=10, n_3=15$} \\ 		
$T$	&0.04	&0.05	&0.07	&0.05	&0.06	&0.07\\
$L$	&0.02	&0.02	&0.04	&0.04	&0.02	&0.04\\
$BL$&0.05	&0.04	&0.06	&0.05	&0.04	&0.06\\
& \multicolumn{6}{c}{$n_1=7, n_2=10, n_3=20$} \\ 	
$T$	&0.04	&0.05	&0.05	&0.07	&0.07	&0.07\\
$L$	&0.02	&0.03	&0.03	&0.04	&0.03	&0.03\\
$BL$&0.05	&0.04	&0.04	&0.05	&0.06	&0.05 \\
& \multicolumn{6}{c}{$n_1=10, n_2=15, n_3=15$} \\ 		
$T$	&0.04	&0.07	&0.06	&0.07	&0.06	&0.07\\
$L$	&0.01	&0.03	&0.03	&0.03	&0.03	&0.05\\
$BL$&0.04	&0.05	&0.05	&0.04	&0.04	&0.06\\
\hline
\end{tabular*}
}
\end{table}

Table 5  displays the simulated power of the tests for equal sample sizes.  The alternative hypothesis has the variance configurations $(\sigma_1^2, \sigma_2^2, \sigma_3^2) = (1,10, 10)$ and the relatively small ratio $(\sigma_1^2, \sigma_2^2, \sigma_3^2) = (1,3, 5)$.  From the same table, it is easily seen that our test $T$ has still the highest power on the average on this array of distributions especially with relatively small sample sizes.  With an average power of $48\%$  for sample size configuration  $n_1=n_2=n_3=7$ and variance ratio $(\sigma_1^2, \sigma_2^2, \sigma_3^2) = (1,10, 10)$, it is $31 \%$ more powerful than \cite{levene60}'s test which is $17\%$, and is $15\%$ higher than the $34\%$ recorded for \cite{limlow96}'s $BL$.  When the sample size is between  7 and 15 (inclusive)  and with variance configuration $(\sigma_1^2, \sigma_2^2, \sigma_3^2) = (1,10, 10)$, our test $T$ is al least $5\%$ more powerful than the Levene's test $L$ and its bootstrap version $BL$.  Similarly, our procedure is $10\%$ more powerful under the variance ratio $(\sigma_1^2, \sigma_2^2, \sigma_3^2) = (1,3, 5)$ and sample size $n_1=n_2=n_3=10$.  Again, the difference in average power (over all the 6 distributions) becomes negligible when the sample sizes exceed $n_1=n_2=n_3=15$ for the variance configuration  $(\sigma_1^2, \sigma_2^2, \sigma_3^2) = (1,10, 10)$.
This strongly suggests that the test $T$ is more sensitive to relatively small departures from homogeneity of the variances than the Levene and the bootstrap Levene tests.

\begin{table}[h!t!b!p!]
\caption{\emph{Estimated power of the tests  at level 0.05 for equal sample sizes. The Monte Carlo estimates are based on 1000 replications, and the standard error of the entries is bounded by 0.016. }} \centerline {
\begin{tabular*}{5.4in}{@{\extracolsep{\fill}}c||c@{\hspace{0.01in}}c@{\hspace{0.01in}}c@{\hspace{0.01in}}c@{\hspace{0.01in}}c@{\hspace{0.01in}}c@{\hspace{0.01in}}|c@{\hspace{0.01in}}}
\hline
Test& Uniform & Normal & Extreme &   Laplace	& Student's $t_5$	& Exponential & Average\\
\hline \hline	
& \multicolumn{6}{c|}{$(\sigma_1^2, \sigma_2^2, \sigma_3^2)=(1, 10, 10)$} \\
\hline 	
& \multicolumn{6}{c|}{$n_1=n_2=n_3=7$} \\ 	
$T$ &0.72	&0.56	&0.48	&0.39	&0.44	&0.32	&0.48\\
$L$	&0.23	&0.19	&0.18	&0.14	&0.14	&0.14	&0.17\\
$BL$&0.48	&0.39	&0.34	&0.29	&0.29	&0.24	&0.34\\
& \multicolumn{6}{c|}{$n_1=n_2=n_3=10$} \\ 	
$T$ & 0.97	&0.85	&0.69	&0.59	&0.67	&0.41	&0.70\\
$L$	&0.74	&0.59	&0.47	&0.37	&0.43	&0.30	&0.48\\
$BL$&0.79	&0.65	&0.51	&0.42	&0.50	&0.34	&0.53\\
& \multicolumn{6}{c|}{$n_1=n_2=n_3=15$} \\ 	
$T$	&1.00	&0.98	&0.85	&0.79	&0.84	&0.58	&0.84\\
$L$	&0.98	&0.90	&0.75	&0.65	&0.74	&0.50	&0.75\\
$BL$&0.98	&0.93	&0.80	&0.69	&0.80	&0.55	&0.79\\
\hline 	
& \multicolumn{6}{c|}{$(\sigma_1^2, \sigma_2^2, \sigma_3^2)=(1,3,5)$} \\
\hline 	
& \multicolumn{6}{c|}{$n_1=n_2=n_3=7$} \\ 		
$T$	&0.28	&0.22	&0.19	&0.19	&0.21	&0.14	&0.20\\	
$L$	&0.09	&0.09	&0.09	&0.07	&0.09	&0.07	&0.08\\	
$BL$&0.27	&0.20	&0.19	&0.16	&0.18	&0.15	&0.19\\	
& \multicolumn{6}{c|}{$n_1=n_2=n_3=10$} \\ 	
$T$	&0.65	&0.44	&0.36	&0.27	&0.35	&0.21	&0.38\\
$L$	&0.38	&0.28	&0.23	&0.17	&0.21	&0.15	&0.24\\
$BL$&0.44	&0.33	&0.28	&0.23	&0.26	&0.18	&0.29\\
& \multicolumn{6}{c|}{$n_1=n_2=n_3=15$} \\ 		
$T$	&0.95	&0.71	&0.50	&0.41	&0.50	&0.30	&0.56\\
$L$	&0.67	&0.50	&0.38	&0.31  &0.38	&0.26	&0.42\\
$BL$&0.76	&0.58	&0.46	&0.37	&0.46	&0.30	&0.49\\
\hline
\end{tabular*}
}
\end{table}

Table 6 demonstrates the performance of the tests when sample sizes are unequal and when the alternative has the small variance ratios  $(\sigma_1^2, \sigma_2^2, \sigma_3^2) = (1,3, 5)$, and  $(5,3,1)$. Just like in the two-sample case, we averaged the power over the two variance ratios. It is clear that the procedure $T$ is the most powerful as indicated by the average of the averaged (over the two small variance ratios) estimated power. Averaging over all the three unequal sample size configurations, the test $T$ has $41\%$  average power. This illustrates that the test $T$  is $24\%$ more powerful than Levene test's  $28\%$, and is $7\%$ more powerful than  the bootstrap version's $35\%$.    Overall, our procedure $T$ still has the least conservative test size estimates and is more powerful in detecting slight departures from the null settings.

\begin{table}[h!t!b!p!]
\caption{\emph{Estimated power at level 0.05 averaged over variance ratios $(\sigma_1^2, \sigma_2^2, \sigma_3^2)= (1, 3, 5)$ and $(5, 3,  1)$  for unequal sample sizes. The Monte Carlo estimates are based on 1000 replications, and the standard error of the entries is bounded by 0.016.}} \centerline {
\begin{tabular*}{5.4in}{@{\extracolsep{\fill}}c||c@{\hspace{0.01in}}c@{\hspace{0.01in}}c@{\hspace{0.01in}}c@{\hspace{0.01in}}c@{\hspace{0.01in}}c@{\hspace{0.01in}}|c@{\hspace{0.01in}}}
\hline
Test& Uniform & Normal & Extreme &   Laplace	& Student's $t_5$	& Exponential & Average\\
\hline \hline
& \multicolumn{6}{c|}{$n_1=7, n_2=10, n_3=15$} \\
$T$ & 0.60	&0.44	&0.32	&0.27	&0.32	&0.22	&0.36\\
$L$ &0.35	&0.28	&0.21	&0.18	&0.21	&0.15	&0.23\\
$BL$ &0.48	&0.37	&0.29	&0.24	&0.28	&0.20	&0.31\\
& \multicolumn{6}{c|}{$n_1=7,  n_2=10, n_3=20$} \\
$T$ &0.64	&0.48	&0.35	&0.31	&0.35	&0.23	&0.40\\
$L$ &0.46	&0.33	&0.27	&0.22	&0.26	&0.20	&0.29\\
$BL$&0.56	&0.41	&0.32	&0.28	&0.32	&0.24	&0.35\\
& \multicolumn{6}{c|}{$n_1=10, n_2=15, n_3=15$} \\
$T$	&0.83	&0.60	&0.43	&0.35	&0.43	&0.24	&0.48\\
$L$ &0.48	&0.37	&0.30  &0.24	&0.28	&0.30	&0.31\\
$BL$&0.61	&0.46	&0.35	&0.28	&0.34	&0.23	&0.38\\
\hline
\end{tabular*}
}
\end{table}

\subsection{Four-Sample Case $(K=3)$}

The estimated levels of the three tests are shown in Table 7 for the 6 sample size configurations $n_1=n_2=n_3=n_4=7, 10, 15$, and $(n_1, n_2, n_3, n_4) = (7, 7, 10,10), (7, 10, 15, 20), (7, 7, 20, 20)$.  The table apparently suggests that  all the three tests  are still robust according to \cite{cojj81}'s criterion for four populations. The test $T$ has a maximum size of 0.07 while the bootstrap Levene's test $BL$  and the Levene's test $L$ have maximum sizes of 0.07 and 0.04, correspondingly. It also shows that our procedure seems to be more conservative than the  bootstrap Levene's test $BL$   under the uniform distribution across all the 5 sample size configurations (except $n_1=n_2=n_3=n_4=15$) or when the sample size is as small as 7.  Mostly, our test $T$ still has the least conservative  Type 1 error estimates among the three procedures for the four-sample case.

\renewcommand{\arraystretch}{1.0}
\begin{table}[h!t!b!p!]
\caption{\emph{Estimated sizes for testing  $H_o:\sigma_1^2=\sigma_2^2=\sigma_3^2=\sigma_4^2$  at level 0.05,  for  different sample size combinations. The Monte Carlo estimates are based on 1000 replications, and the standard error of the entries is bounded by 0.016.}} \centerline {
\begin{tabular*}{5.4in}{@{\extracolsep{\fill}}c||c@{\hspace{0.01in}}c@{\hspace{0.01in}}c@{\hspace{0.01in}}c@{\hspace{0.01in}}c@{\hspace{0.01in}}c@{\hspace{0.01in}}}
\hline
Test & Uniform & Normal & Extreme &   Laplace	& Student's $t_5$	& Exponential\\
\hline \hline
& \multicolumn{6}{c}{$n_1=n_2=n_3=n_4=7$} \\ 		
$T$ &0.02	&0.03	&0.03	&0.04	&0.03	&0.04\\
$L$	&0.01	&0.01	&0.01	&0.02	&0.02	&0.03\\
$BL$&0.02	&0.04	&0.06	&0.05	&0.06	&0.07\\
& \multicolumn{6}{c}{$n_1=n_2=n_3=n_4=10$} \\
$T$ & 0.03	&0.05	&0.06	&0.05	&0.06	&0.05\\
$L$	&0.02	&0.03	&0.04	&0.03	&0.03	&0.04\\
$BL$&0.04	&0.05	&0.05	&0.05	&0.06	&0.06\\
& \multicolumn{6}{c}{$n_1=n_2=n_3=n_4=15$} \\
$T$	&0.05	&0.06	&0.06	&0.06	&0.05	&0.06\\
$L$	&0.02	&0.03	&0.03	&0.04	&0.02	&0.04\\
$BL$&0.04	&0.05	&0.04	&0.06	&0.03	&0.05\\
\hline
& \multicolumn{6}{c}{$n_1=7, n_2=7, n_3=10, n_4=10$} \\		
$T$	&0.02	&0.04	&0.03	&0.04	&0.04	&0.06\\	
$L$	&0.01	&0.02	&0.03	&0.03	&0.03	&0.03\\	
$BL$&0.04	&0.05	&0.05	&0.05	&0.04	&0.06\\	
& \multicolumn{6}{c}{$n_1=7, n_2=10, n_3=15, n_4=20$} \\
$T$	&0.04	&0.05	&0.04	&0.07	&0.05	&0.06\\
$L$	&0.02	&0.03	&0.03	&0.03	&0.02	&0.03\\
$BL$&0.04	&0.04	&0.05	&0.05	&0.04	&0.05\\
& \multicolumn{6}{c}{$n_1=7, n_2=7, n_3=20, n_4=20$} \\	
$T$	&0.03	&0.05	&0.04	&0.07	&0.05	&0.04\\
$L$	&0.03	&0.02	&0.02	&0.03	&0.03	&0.03\\
$BL$&0.06	&0.04	&0.05	&0.05	&0.05	&0.05\\
\hline
\end{tabular*}
}
\end{table}

Tables 8 and 9 give the estimated power of the tests when the sample sizes are equal and unequal, respectively.  The power of the tests is computed using four variance and six sample size combinations. When the sample sizes are equal, we considered the variance ratios $(\sigma_1^2, \sigma_2^2, , \sigma_3^2, \sigma_4^2) = (1,10, 10, 10)$,  and $(1,16,11, 16)$ for the alternative.  Table 8 below shows that the test $T$ is at least $12\%$ more powerful than the other tests when $n_i=10$ under the two variance ratios. When the alternative assumes the variance ratio $(\sigma_1^2, \sigma_2^2,  \sigma_3^2, \sigma_4^2) = (1,10, 10, 10)$ and sample size $n_i=15$, our test $T$ appeared to be more powerful for populations with high kurtosis (e.g., exponential). In addition, a direct comparison of  our results with that of  Lim and Loh (1996)'s corresponding to the variance configuration $(\sigma_1^2, \sigma_2^2, , \sigma_3^2, \sigma_4^2) = (1,6, 11, 16)$ and sample sizes $n_i=10$ indicates that the proposed test is more powerful than the  bootstrap Bartlett's test (except the  exponential distribution). These results further validate the superiority of our test when the sample sizes are equal.

\begin{table}[h!t!b!p!]
\caption{\emph{Estimated power of the tests  at level 0.05 for variance ratios $(\sigma_1^2, \sigma_2^2, \sigma_3^2, \sigma_4^2)=(1, 10, 10, 10)$ and $(1, 6, 11, 16)$,  and for  equal sample sizes. The Monte Carlo estimates are based on 1000 replications, and the standard error of the entries is bounded by 0.016. }} \centerline {
\begin{tabular*}{5.4in}{@{\extracolsep{\fill}}c||c@{\hspace{0.01in}}c@{\hspace{0.01in}}c@{\hspace{0.01in}}c@{\hspace{0.01in}}c@{\hspace{0.01in}}c@{\hspace{0.01in}}|c@{\hspace{0.01in}}}
\hline
Test& Uniform & Normal & Extreme &   Laplace	& Student's $t_5$	& Exponential & Average\\
\hline \hline
& \multicolumn{6}{c|}{$(\sigma_1^2, \sigma_2^2, \sigma_3^2, \sigma_4^2)=(1, 10, 10, 10)$} \\	
\hline
& \multicolumn{6}{c|}{$n_1=n_2=n_3=n_4=7$} \\
$T$ &0.63	&0.52	&0.40	&0.33	&0.39	&0.25	&0.42\\
$L$	&0.12	&0.15	&0.11	&0.10	&0.10	&0.09	&0.11\\
$BL$&0.38	&0.33	&0.25	&0.22	&0.23	&0.19	&0.27\\
& \multicolumn{6}{c|}{$n_1=n_2=n_3=n_4=10$} \\
$T$ & 0.98	&0.85	&0.65	&0.56	&0.66	&0.37	&0.68\\
$L$	&0.64	&0.47	&0.36	&0.27	&0.35	&0.25	&0.39\\
$BL$&0.72	&0.55	&0.41	&0.32	&0.43	&0.28	&0.45\\
& \multicolumn{6}{c|}{$n_1=n_2=n_3=n_4=15$} \\
$T$	&1.00	&0.99	&0.84	&0.77	&0.82	&0.56	&0.83\\
$L$	&0.97	&0.84	&0.67	&0.55	&0.66	&0.40	&0.68\\
$BL$&0.99	&0.90	&0.73	&0.62	&0.74	&0.44	&0.74\\
\hline
& \multicolumn{6}{c|}{$(\sigma_1^2, \sigma_2^2, \sigma_3^2, \sigma_4^2)=(1, 6, 11, 16)$} \\
\hline
& \multicolumn{6}{c|}{$n_1=n_2=n_3=n_4=7$} \\
$T$	&0.62	&0.52	&0.40	&0.34	&0.40	&0.27	&0.42\\
$L$	&0.24	&0.20	&0.18	&0.17	&0.18	&0.14	&0.19\\
$BL$&0.50	&0.42	&0.34	&0.31	&0.33	&0.24	&0.36\\
& \multicolumn{6}{c|}{$n_1=n_2=n_3=n_4=10$} \\
$T$	&0.97	&0.86	&0.65	&0.58	&0.67	&0.39	&0.69\\
$L$	&0.79	&0.61	&0.47	&0.41	&0.48	&0.32	&0.51\\
$BL$&0.83	&0.67	&0.53	&0.45	&0.55	&0.36	&0.56\\
& \multicolumn{6}{c|}{$n_1=n_2=n_3=n_4=15$} \\
$T$	&1.00	&0.99	&0.85	&0.79	&0.85	&0.57	&0.84\\
$L$	&0.98	&0.94	&0.81	&0.68	&0.78	&0.53	&0.79\\
$BL$&0.99	&0.96	&0.85	&0.74	&0.84	&0.59	&0.83\\
\hline
\end{tabular*}
}
\end{table}

Table 9 reflects the performance of the three tests when the sample sizes are unequal. We compared the power using the variance configurations  $(\sigma_1^2, \sigma_2^2, \sigma_3^2, \sigma_4^2)= (1, 6, 11, 16)$ and $(16, 11, 6,  1)$.  We also considered three sample size configurations $(n_1,  n_2, n_3,  n_4) = (7,7,10,10), (7,7,20,20)$, and $(7,10,15,20)$ and averaged the power over the two variance ratios. Among the three tests considered, our test $T$ still is more powerful even when the sample sizes are not equal. It has an overall average power of $67\%$ while Levene's $L$ and the bootstrap test $BL$ have $53\%$ and $60\%$ power, correspondingly.

Based on these simulation results, we observed that our procedure $T$   has the least conservative test size estimates and is more powerful than the Levene's test $L$ and its bootstrap version $BL$.  These results also confirmed that bootstrapping Levene's test  $L$ generally improved the Type I and Type II errors in most cases.

\begin{table}[h!t!b!p!]
\caption{\emph{Estimated power at level 0.05 averaged over variance ratios $(\sigma_1^2, \sigma_2^2, \sigma_3^2, \sigma_4^2)= (1, 6, 11, 16)$ and $(16, 11, 6,  1)$  for unequal sample sizes. The Monte Carlo estimates are based on 1000 replications, and the standard error of the entries is bounded by 0.016.}} \centerline {
\begin{tabular*}{5.4in}{@{\extracolsep{\fill}}c||c@{\hspace{0.01in}}c@{\hspace{0.01in}}c@{\hspace{0.01in}}c@{\hspace{0.01in}}c@{\hspace{0.01in}}c@{\hspace{0.01in}}|c@{\hspace{0.01in}}}
\hline
Test& Uniform & Normal & Extreme &   Laplace	& Student's $t_5$	& Exponential & Average\\
\hline \hline
& \multicolumn{6}{c|}{$n_1=7,  n_2=7, n_3=10,  n_4=10$} \\
$T$	&0.84	&0.71	&0.54	&0.44	&0.52	&0.30	&0.56\\
$L$	&0.53	&0.43	&0.33	&0.26	&0.32	&0.22	&0.35\\
$BL$&0.68	&0.56	&0.44	&0.35	&0.42	&0.28	&0.46\\
& \multicolumn{6}{c|}{$n_1=7, n_3=10,  n_3=15, n_4=20$} \\
$T$ &0.97	&0.89	&0.73	&0.66	&0.73	&0.47	&0.74\\
$L$ &0.90	&0.73	&0.59	&0.51	&0.60	&0.43	&0.63\\
$BL$ &0.93	&0.80	&0.67	&0.57	&0.66	&0.47	&0.68\\
& \multicolumn{6}{c|}{$n_1=7, n_2=7, n_3=20,  n_4=20$} \\
$T$ &0.94	&0.87	&0.72	&0.63	&0.72	&0.44	&0.72\\
$L$ &0.88	&0.74	&0.60	&0.47	&0.57	&0.38	&0.60\\
$BL$&0.93	&0.81	&0.66	&0.55	&0.64	&0.43	&0.67\\
\hline
\end{tabular*}
}
\end{table}

\newpage
\section{Concluding Remarks} 

We have proposed a  variance-based statistic that led to a bootstrap test for heterogeneity of variances.  Our procedure, which used a box-type  acceptance region is shown to be more  sensitive to slight deviations from the null specifications. The tests were compared using a considerable number of sample-size, variance-ratio, and number-of-population combinations. We have also used a random number generator that has desirable properties. The confidence region approach  showed some promising results especially for the two-sample $(K=1)$ case. It has the potential of multiplying the power of Levene's test $L$. It is also observed that slight departures from the null required larger sample sizes (and are preferably equal) to achieve good power. Overall, simulation results indicated that our test $T$ is  more powerful compared with the Levene's test $L$ and \cite{limlow96}'s procedure $BL$  and is mostly the least  conservative  procedure in controlling the Type I error rate.

Furthermore, our results shared similar caveats with that of \cite{limlow96}. For instance, the properties of our test may change when there are more than four populations involved, and these populations are not from a location-scale family and may have different kurtosis. This means that experimenters should exercise caution when our method is used in practice. Within the boundaries of our study, we generally recommend the test $T$ under most conditions. However, we recommend \cite{limlow96}'s procedure $BL$ for small samples $(n_i<7)$. 

Finally, we would like to extend our study to evaluating the performance of these tests for more leptokurtic distributions. We also wish to construct a rectangular prism  with unequal lengths or a non-box-type acceptance region as in \cite{hal87}. Employing better bootstrap techniques (e.g., variance stabilization as in \cite{tib88}, pooling residuals as in \cite{bbrow89},  bias correction,  balance  and  weighted bootstraps) to enhance the performance of our test would be of interest as well.  A more efficient procedure in calculating  the critical value  involving larger bootstrap samples $B$ and simulation runs would be desirable.  Comparing the proposed method with the bootstrap version of Bartlett's test   \citep[see][]{bbrow89,limlow96} especially for larger numbers of groups or populations $(K=16, say)$   would also be worth pursuing.